\documentclass[12pt]{article}
\usepackage{graphicx}

\begin{document}

\title{\bf The dynamical temperature and the standard map}

\author{I.~I.~Shevchenko\/\thanks{E-mail:~iis@gao.spb.ru} \\
Pulkovo Observatory of the Russian Academy of Sciences \\
Pulkovskoje ave.~65/1, St.Petersburg 196140, Russia}
\date{}

\maketitle


\begin{center}
Abstract
\end{center}

\noindent Numerical experiments with the standard map at high
values of the stochasticity parameter reveal the existence of
simple analytical relations connecting the volume and the dynamical
temperature of the chaotic component of the phase space.

\bigskip


\section{Introduction}

The standard map is an important object in studies in nonlinear
dynamics, mainly because it is often used to describe the local
behavior of other more complicated symplectic
maps~\cite{C78,C79,LL92}. What is more, it serves as an important
independent mechanical and physical paradigm~\cite{C96,M92}.
General properties of the standard map, often in interrelation
with those of the separatrix map, Fermi map and other fundamental
maps, were considered and studied in detail
in~\cite{C78}--\cite{VC98}, and many other works.

The standard map is given by the equations

\begin{eqnarray}
     y_{i+1} &=& y_i + \frac{K}{2 \pi} \sin (2 \pi x_i) \ \ \ (\mbox{mod } 1), \nonumber \\
     x_{i+1} &=& x_i + y_{i+1} \ \ \ (\mbox{mod } 1),
\label{stm}
\end{eqnarray}

\noindent where $K$ is the so-called stochasticity
parameter~\cite{C78,C79}.

In Ref.~\cite{S04c} we studied three major characteristics of the
chaotic dynamics of the standard map, namely, the measure $\mu$ of
the main connected chaotic domain (MCCD), the maximum Lyapunov
exponent $L$ of the motion in this domain, and the dynamical
entropy $h = \mu L$, as functions of the stochasticity parameter
$K$. The perturbations of the domain due to the birth and
disintegration of islands of stability, upon variations of $K$,
were considered in particular. By means of extensive numerical
experiments, we showed that these perturbations are isentropic, at
least approximately: the dynamical entropy follows the smooth
dependence $h(K) = \ln {K \over 2} + {1 \over K^2}$, and does not
fluctuate about it with changing the stochasticity parameter,
while local jumps in $\mu$ and $L$ due to the birth and
disintegration of regular islands are significant.

On the other hand, recently Baldovin~\cite{B04} and Baldovin et
al.~\cite{BBT04} introduced the notion of the ``dynamical
temperature'' for the standard map, defining it as the variance of
the momentum:

\begin{equation}
T \equiv \langle (y - \langle y \rangle )^2 \rangle - \frac{1}{12}
= \langle y^2 \rangle - \langle y \rangle^2 - \frac{1}{12},
\label{TB}
\end{equation}

\noindent where the angle brackets signify ensemble average; the
constant addend is the normalizing shift, introduced for
convenience. This choice of normalization gives $T=0$ for the case
of the uniform ensemble~\cite{B04,BBT04}. Henceforth we use
different normalization:

\begin{equation}
T \equiv 12 \langle (y - \langle y \rangle )^2 \rangle = 12
(\langle y^2 \rangle - \langle y \rangle^2).
\label{T}
\end{equation}

\noindent This gives $T=1$ for the case of the uniform ensemble.
This normalization is used in calculating $T$ in what follows. In
practice, the measurement of $T$ is performed by averaging over an
iterated trajectory in the chaotic domain, i.e., we assume that
the ensemble average is equal to the time average. This is
justified in the considered situation, when the regular islands
are small and the motion is almost completely ergodic on the phase
plane. The relation between ensemble and time averages for
low-dimensional symplectic maps in the opposite case of extended
borders between chaotic and regular components of the phase space
were considered and analyzed in Ref.~\cite{B04}.

In this paper, we investigate the behavior of two state variables
of the MCCD of the phase space of the standard map: namely, the
volume $\mu$ and the dynamical temperature $T$, both as functions
of the stochasticity parameter $K$. At moderate values of $K$ (at
approximately $K < 4$), the non-monotonic (spike) variations in
$\mu(K)$ and $T(K)$ are conditioned by the process of absorption
of minor chaotic domains by the MCCD, while $K$ increases; at
larger values of $K$ (at approximately $K > 4$), they are
conditioned by the process of birth and disintegration of
stability islands. In other words, the perturbations of the
chaotic domain at high values of $K$ are due to the birth and
disintegration of islands of stability, upon small variations of
$K$. By means of extensive numerical experiments, we show that the
variations of $\mu$ and $T$ due to this process obey simple
analytical relations.

\section{Numerical experiments}

The traditional ``one trajectory method''
(OTM)~\cite{C78,C79,SM03,S04c} has been used for calculation of
$\mu$. This method consists in computing the number of cells
explored by a single trajectory on a grid exposed on the phase
plane. A strict but computationally much more expensive approach
for measuring $\mu$ consists in calculating the values of the
coarse-grained area of the chaotic component for a set of various
resolutions of the grid, in order to find the asymptotic value of
$\mu$ at the infinitely fine resolution (see~\cite{UF85}). In some
cases the ``LCE segregation method''
(LCESM)~\cite{C78,C79,SM03,S04c} has been employed for
verification of the obtained values of $\mu$. In this method, a
numeric criterion is used for separation of the regular and
chaotic trajectories. The criterion is provided by an analysis of
the differential distribution of the values of the finite-time
Lyapunov characteristic exponents (LCEs) computed on a set of
trajectories with the starting values randomly generated on the
phase plane or specified on a regular grid on the phase plane. In
the distribution, the peak corresponding to the regular
trajectories is movable: if one increases the LCE computation
time, the peak moves to the left on the abscissa axis, because the
computed LCEs for the regular trajectories tend to zero; while all
the peaks corresponding to the chaotic domains stay immovable, on
condition that the computation time is long enough. By increasing
the computation time one can make the movable and immovable peaks
completely distinct, and in this way determine the ``finite-time
LCE'' value separating the regular and chaotic trajectories for
the given computation time. The OTM and LCESM were both proposed
and used by Chirikov~\cite{C78,C79} in computations of $\mu$ for
the standard map. Analogous methods were used in Ref.~\cite{SM03}
in computations of the chaotic domain measure in the
H\'enon--Heiles problem, and in Ref.~\cite{S04c} for computations
of $\mu$ for the standard map. A detailed description of the
currently used versions of the methods is given in
Ref.~\cite{S04c}.

In Figs.~1--4, we present the obtained numerical data on the
measure $\mu$ and the dynamical temperature $T$ of the MCCD, and
their interrelations. Fig.~1 shows $\mu(K)$ and $T(K)$ for a
perturbation of the MCCD due to the birth and disintegration of
the regular islands corresponding to a 2-periodic solution (at $K
\in (11.9, 12.6)$) and to an accelerator mode (at $K \in (12.6,
13.3)$). An accelerator mode represents a 1-periodic solution on
the torus $x, y \in [0, 1] \times [0, 1]$. Theoretical ranges in
$K$ for the existence of both types of solutions are given
in~\cite{C79}; the narrow ``windows'' of their existence follow
with the period of $2 \pi$ in $K$, and with increasing $K$ they
become more and more narrow. In Fig.~2, the same plots are
constructed for a perturbation conditioned by the birth and decay
of the islands due to a 4-periodic solution at $K \approx 9.21
\div 9.45$.

The OTM has been used for calculating $\mu$; the grid is $5000
\times 5000$ pixels on the square $x, y \in [0, 1] \times [0, 1]$.
The map has been iterated $n_{it} = 10^9$ times at each value of
$K$. The step in $K$ is equal to $0.005$ (Fig.~1) and $0.001$
(Fig.~2). Each value of $T$ has been computed simultaneously with
the value of $\mu$ for the same trajectory.

No error bars are shown in the Figs.~1 and 2, since the typical
errors appear to be rather low. One can estimate them in the
following way. Consider first the calculation of $\mu$. The main
uncertainty in the measured $\mu$ value, at sufficiently long
iteration times $n_{it}$, is conditioned by the resolution of the
grid imposed on the phase plane. We designate the number of pixels
at each side of the square $x, y \in [0, 1] \times [0, 1]$ by $l$;
so, $l = 5000$. Consider the model case of a round islet (the
choice of geometries is not principal for our estimate by the
order of magnitude). The relative uncertainty $\varepsilon$ in the
measured area of the disk is given approximately by one half of
the ratio of the area of the chain of the pixels at the disk
border and the disk area $S$. It is straightforward to show that
$\varepsilon = l^{-1} (\pi / S)^{1/2}$. This is the relative error
by which the island area can be overestimated by the OTM. For the
accelerator modes the maximum area of one island scales as $0.19
K^{-2}$~\cite{C96}; so, one has $\varepsilon \approx 4 K l^{-1}$.
For the accelerator mode in Fig.~1 this formula gives $\varepsilon
\approx 0.010$. This is the relative uncertainty in the measure
$\mu_{reg}$ of the regular component of the phase space. Hence the
absolute uncertainty in $\mu$ (in the ordinate in Fig.~1) is equal
to $2 \cdot 10^{-5}$; i.e., the error is smaller than the size of
the data point symbols in the Figure. Practically the same error
values are derived in the case of the plot in Fig.~2, i.e., in the
case of the 4-periodic solution.

In what concerns the errors in calculation of $T$, they seem to be
controlled only by the computation time $n_{it}$. If one decreases
this time, the ``noise'' appears in the plot: instead of the
smooth long-range (in comparison with the step in $K$) variations,
$T$ varies irregularly at each step in $K$; the amplitude of these
irregular one-step variations gives the typical error in the
estimation of $T$. From the fact of the absence of such variations
in Figs.~1 and 2, one concludes that the errors are typically
smaller than at least one-half of the distance in $\mu$ between
each two consecutive points on the plots, i.e., they are
insignificant in the given case of $n_{it} = 10^9$.

\section{Analytical relations between volume and dynamical
temperature}

The local minima in $\mu(K)$ seen in Figs.~1 and 2 clearly
correspond to local extrema in $T(K)$. The observed amplitudes of
the volume and temperature variations are apparently
commensurable, namely, their ratios represent some integer or
half-integer numbers. This suggests the existence of the power law
relations with rational exponents. We verify this hypothesis by
constructing the ``$\mu$--$T$'' graphs in Figs.~3 and 4. All the
points present in Figs.~1 and 2 are depicted in Figs.~3 and 4; in
the latter Figures, each pair of neighbors in $K$ are connected by
straight segments.

For all $K > 5$ the islands are small, i.e., $\mu$ is close to 1;
this is discussed in detail at the end of this Section. In this
context, it clearly follows from Figs.~3 and 4 that the relation

\begin{equation}
T = 1 - \gamma (1 - \mu)
\label{Tmu}
\end{equation}

\noindent takes place with three different values of the $\gamma$
constant. Namely, $\gamma = -1$ in the case of the 2-periodic
solution, $\gamma = 2$ for the accelerator mode, and $\gamma =
1/2$ for the 4-periodic solution.

Among 320 data points present in Fig.~3 there are 10 (i.e.,
$\approx 3$\%) that notably (by more than $0.0005$) deviate from
the dotted lines given by Eq.~(\ref{Tmu}) with $\gamma = -1$ and
$2$; among 350 data points in Fig.~4 there are 8 (i.e., $\approx
2$\%) that notably (by more than $0.0002$) deviate from the dotted
line given by Eq.~(\ref{Tmu}) with $\gamma = 1/2$. In other words,
the percentage of notable deviations is rather low. The deviating
points apparently correspond to those values of $K$ at which
secondary resonance chains separate from the outer borders of the
islands; these phenomena manifest themselves in the secondary
peaks in the curves $\mu(K)$ and $T(K)$. In such situations the
measuring of both $\mu$ and $T$ are subject to additional
uncertainties: the thin chaotic layers between the islands and the
surrounding resonance chains can be unresolved in the adopted grid
of pixels, and, besides, the enhanced sporadic sticking of the
chaotic trajectories to the islands may lead to increasing the
contribution of the ``temperature on the islands'' (this notion is
considered below) in the total measured temperature value.

Apart from these casual deviations, the sticking phenomenon does
not influence the temperature measurements. (This is contrary to
the case of the diffusion coefficient, which, as it is well known,
diverges in the presence of the accelerator modes,
see~\cite{C79,C96}.) If a trajectory constantly sticks to an
accelerator mode island, the temperature tends to the
``temperature on the island'' value, which is finite, because the
temperature~(\ref{TB}) and (\ref{T}) is defined in the factored
phase space: $y$ is taken modulo 1.

If one puts Eq.~(\ref{Tmu}) in the form $\frac{\Delta T}{T} =
\gamma \frac{\Delta \mu}{\mu}$ and integrates, one has

\begin{equation}
T = \mu^\gamma.
\label{plgamma}
\end{equation}

\noindent Eqs.~(\ref{Tmu}) and (\ref{plgamma}) are practically
equivalent for any application. At $K=9.24$, the MCCD area $\mu
\approx 0.995$; and the values of $T$ given by Eqs.~(\ref{Tmu})
and (\ref{plgamma}) at this value of $\mu$ differ by only $2.5
\cdot 10^{-5}$. For smaller islands, emerging at greater values of
$K$, the difference is smaller.

We have considered the behavior of the standard map at two
definite intervals of $K$. Now let us look at the problem in a
global perspective. In Fig.~5, the dependences $\mu(K)$ and $T(K)$
are presented in a broad range: $K \in [4, 30]$. The values of
$\mu(K)$ and $T(K)$ are calculated as previously, except that a
lower grid resolution ($2000 \times 2000$ pixels) and shorter
trajectories ($n_{it} = 10^8$) are used. The step in $K$ is equal
to $0.01$. The observed $\mu(K)$ dependence (Fig.~5) at large
values of $K$, i.e., at $K$ approximately greater than $6$,
represents the basic line $\mu = 1$ interrupted by periodic
sequences of sharp narrow minima whose depth diminishes with $K$.
The most prominent of these sequences correspond to the 2-periodic
solutions and accelerator modes (at $K_m \approx 2 \pi m$; $m=1,
2, \ldots$) and the 4-periodic solutions (at $K_m \approx 2 \pi
\left(m + \frac{1}{2} \right)$; $m=1, 2, \ldots$). The less
pronounced sequences of minima correspond to periodic solutions of
higher order. The observed minimum values of $\mu$ for these three
basic cases of periodic solutions agree with the semi-analytical
scaling $\mu_{reg} = 0.38 K^{-2}$ derived by Chirikov~\cite{C96}
for the maximum area of the accelerator mode islands. The
difference is within $(1 \div 2) \cdot 10^{-4}$. This agreement
verifies the good accuracy of our measurements of $\mu$. Note that
recently Giorgilli and Lazutkin~\cite{GL00} rigorously derived the
inversely quadratic law of decay with $K$ of the area of the main
regular islands, i.e., the islands emerging at $K \approx \pi n$
($n=2, 3, \ldots$).

The variations with $K$ in the standard map behavior due to the
2-periodic solutions and accelerator modes repeat themselves at
$m=2, 3, \dots$, only the amplitude of the variations diminish.
The same is true for the 4-periodic solutions. So, it is natural
to expect that the laws~(\ref{Tmu}) and (\ref{plgamma}) are
universal for all $m$. We have performed recomputation of our
plots taking $K \in [18.4, 19.3]$ (a 2-periodic solution and an
accelerator mode, $m=3$) and $K \in [15.58, 15.67]$ (a 4-periodic
solution, $m=2$). As expected, the values of $\gamma$ have been
found to reproduce the cases analyzed above.

\section{Theoretical explanation}

Finally, we consider a tentative explanation for the observed
relations~(\ref{Tmu}) and (\ref{plgamma}). In Ref.~\cite{S04c}, we
showed that the dynamical entropy  $h = \mu L$, where $L$ is the
maximum Lyapunov exponent of the motion in the MCCD, follows the
smooth dependence $h(K) = \ln {K \over 2} + {1 \over K^2}$, and
does not fluctuate about it with changing $K$, while local jumps
in $\mu$ and $L$ due to the birth and disintegration of regular
islands are significant. In other words, the $h(K)$ function
normalized by the ``long-term'' trend $\ln {K \over 2} + {1 \over
K^2}$ is constant and thus represents an invariant with respect to
$K$. Could a similar invariant be found in the case of the
dynamical temperature? Note that the dynamical entropy  $h = \mu
L$ is nothing but the maximum Lyapunov exponent averaged over the
whole phase space, because the maximum Lyapunov exponent is zero
on the islands. Since the dynamical temperature is, generally
speaking, non-zero on the islands, an analogous temperature
invariant, which is the dynamical temperature averaged over the
whole phase space including the chaotic sea and the islands, would
have the form $\Theta = \mu_{sea} T_{sea} + \mu_{isles} T_{isles}
= 1$ in our normalization of $T$. For the sake of clarity we
designate $\mu_{sea}\equiv\mu$, $T_{sea}\equiv T$; $\mu_{isles}$,
equal to $1-\mu$, is the measure of the islands; $T_{isles}$ is
the temperature on the islands, i.e., the quantity~(\ref{T})
calculated for the initial data on the islands. From this equality
one has

\begin{equation}
T = \frac{1 - \mu_{isles} T_{isles}}{1 - \mu_{isles}} = 1 +
\mu_{isles} (1 - T_{isles}) + O(\mu_{isles}^2) .
\label{Tth}
\end{equation}

\noindent Comparing Eqs.~(\ref{Tmu}) and (\ref{Tth}) with the
precision of $O(\mu_{isles}^2)$, we find

\begin{equation}
\gamma \approx T_{isles} - 1.
\label{gm}
\end{equation}

\noindent The quantity $T_{isles}$ is determined by the $y$
location of the islands in the phase space. The data on the
location of the islands is available in~\cite{C79} (for the
2-periodic solutions and accelerator modes) and in~\cite{C78} (for
the 4-periodic solutions). The temperature is straightforwardly
calculated by Eq.~(\ref{T}), and it has the values: $T_{isles}
\approx 0$, $3$, and $3/2$ for the 2-periodic solutions,
accelerator modes, and 4-periodic solutions, respectively; i.e.,
$\gamma \approx -1$, $2$, and $1/2$, respectively, in accord with
our analysis of the plots in Figs.~3 and 4.

\section{Conclusions}

Let us summarize the main results. The measure $\mu$ of the main
connected chaotic domain (MCCD) of the standard map and its
dynamical temperature $T$ have been studied as functions of the
stochasticity parameter $K$ by means of extensive numerical
experiments. The process of the birth and disintegration (upon
small variations of $K$) of the islands of stability inside the
chaotic domain result in local variations (spikes) in $\mu(K)$ and
$T(K)$. It has been shown that the variations of $\mu$ and $T$ are
connected by certain dependences~(\ref{Tmu}). Since the island
areas are small, these dependences are equivalent to power
laws~(\ref{plgamma}): $T = \mu^\gamma$, with $\gamma = -1$ and $2$
for the 2-periodic solutions and accelerator modes respectively
and $\gamma = 1/2$ for the 4-periodic solutions. It is interesting
to note that the value of $\gamma$ for the 4-periodic solutions is
the arithmetic mean of the values for the 2-periodic solutions and
accelerator modes.

If the newly found relations are confirmed analytically in a
rigorous way, they would provide an efficient instrument for
estimating the measure of chaotic domains by means of
computing the dynamical temperature. Such a method would be more
efficient than the OTM or LCESM, since any direct measurement of
$\mu$ is a tedious and time-consuming task, subject to a number of
uncertainties.

The author is thankful to anonymous referees for advice and
remarks which led to significant improvement of the manuscript.
This work was supported by the Russian Foundation for Basic
Research (project \# 05-02-17555) and by the Programme of
Fundamental Research of the Russian Academy of Sciences
``Fundamental Problems in Nonlinear Dynamics''. The computations
were partially carried out at the St.\ Petersburg Branch of the
Joint Supercomputer Centre of the Russian Academy of Sciences.

\newpage

\begin{figure}
\centering
\includegraphics[width=0.75\textwidth]{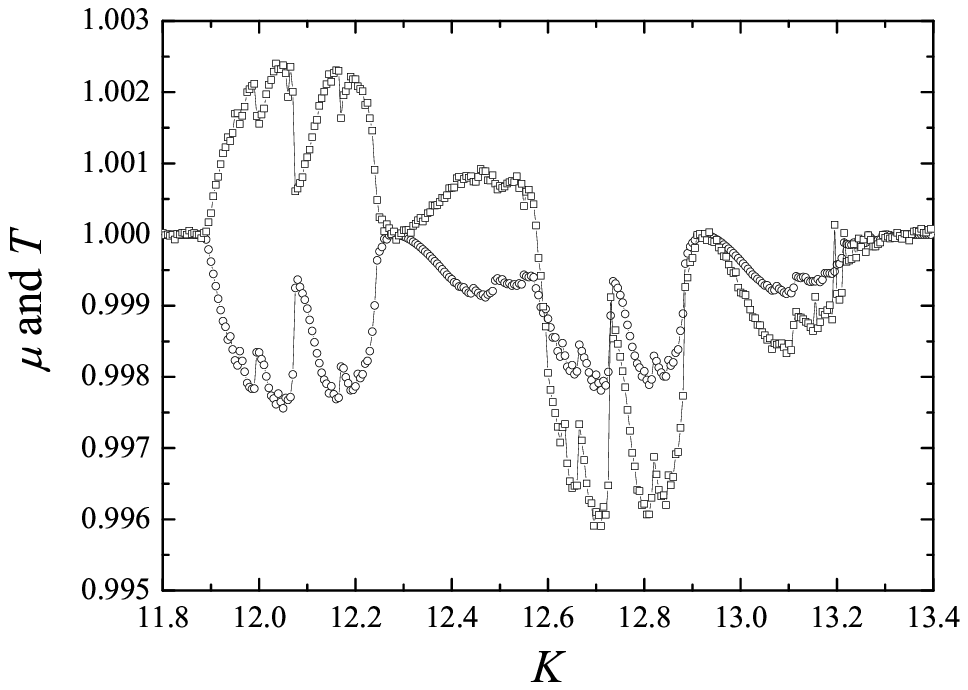}
\caption{The volume $\mu$ (circles) and the dynamical temperature
$T$ (squares) of the MCCD, in function of $K$, for a perturbation
due to a 2-periodic solution and an accelerator mode.}
\label{fig1}
\end{figure}

\begin{figure}
\centering
\includegraphics[width=0.75\textwidth]{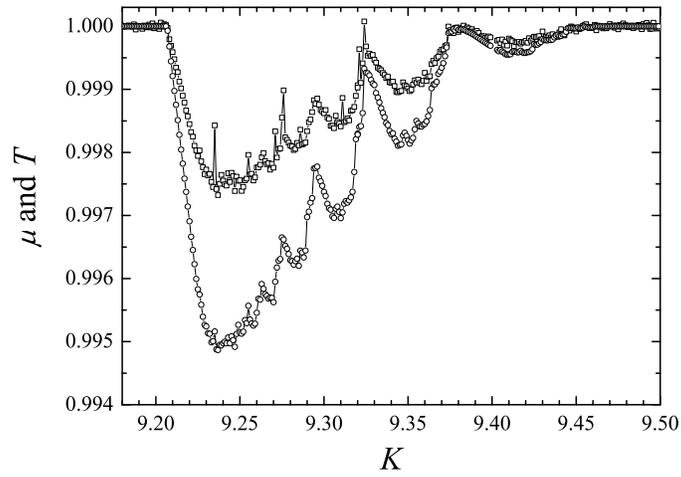}
\caption{The same for a perturbation due to a 4-periodic
solution.}
\label{fig2}
\end{figure}

\begin{figure}
\centering
\includegraphics[width=0.75\textwidth]{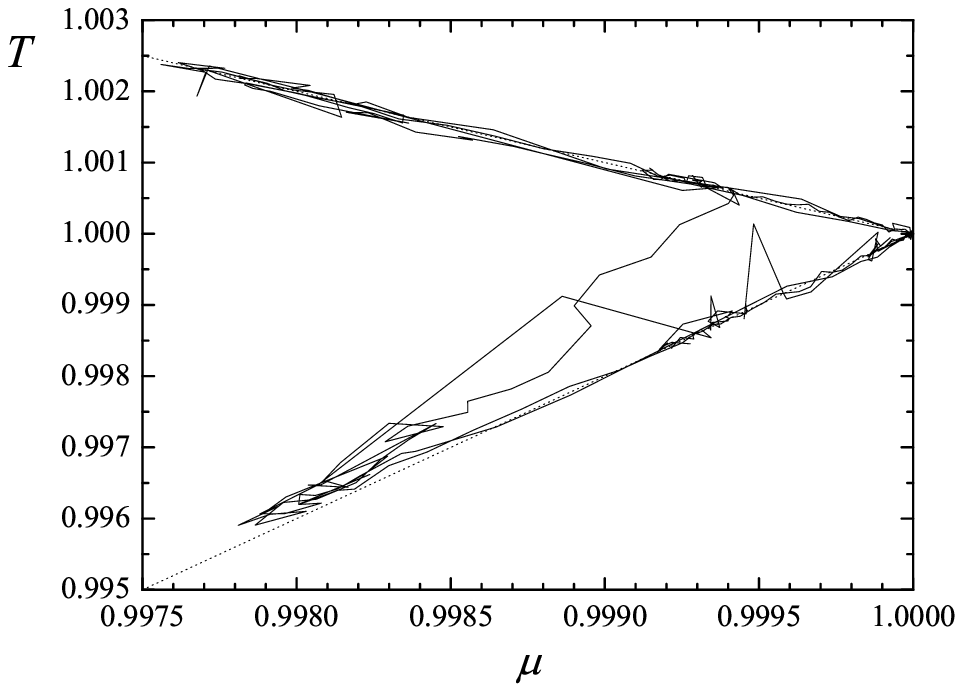}
\caption{The ``$\mu$--$T$'' relation (solid polygonal line) for
the perturbation shown in Fig.~1. Relations~(\ref{plgamma}) with
$\gamma = -1$ and $2$ are given by the dotted straight lines.}
\label{fig3}
\end{figure}

\begin{figure}
\centering
\includegraphics[width=0.75\textwidth]{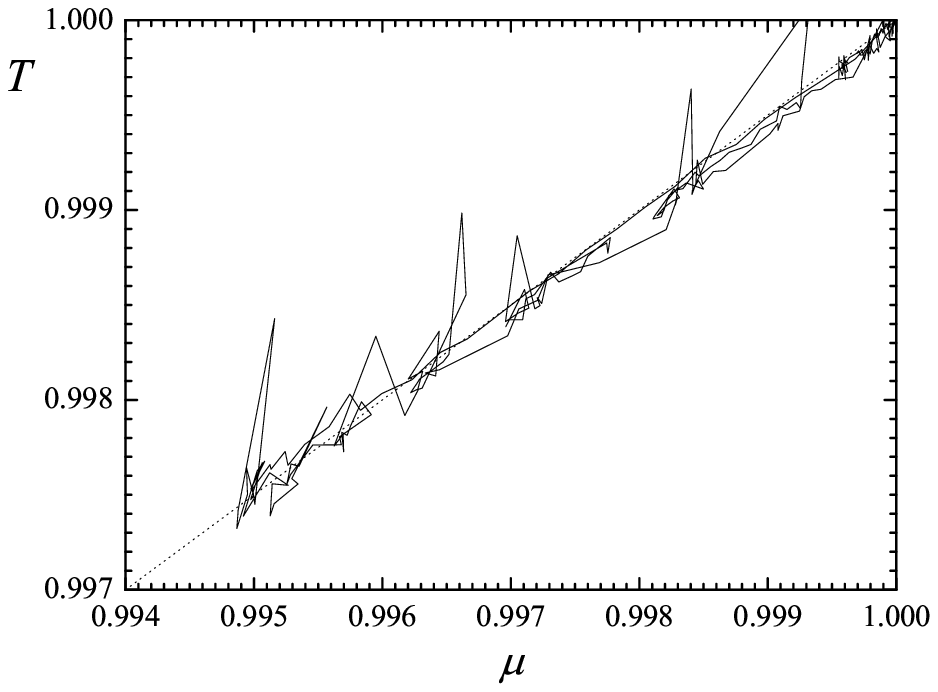}
\caption{The ``$\mu$--$T$'' relation (solid polygonal line) for
the perturbation shown in Fig.~2. Relation~(\ref{plgamma}) with
$\gamma = 1/2$ is given by the dotted straight line.}
\label{fig4}
\end{figure}

\begin{figure}
\centering
\includegraphics[width=0.75\textwidth]{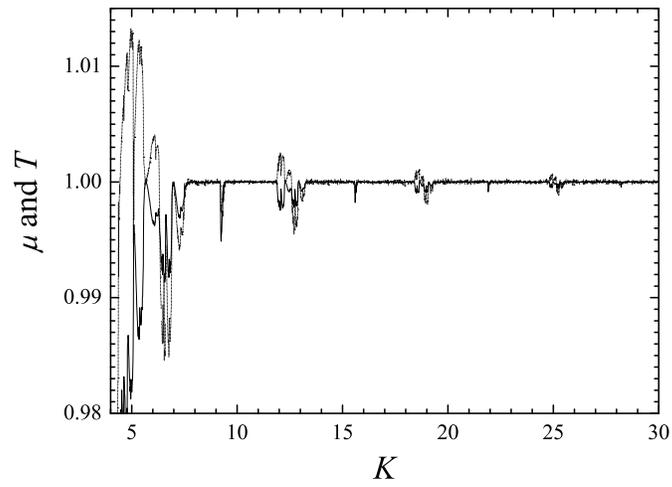}
\caption{The $\mu(K)$ and $T(K)$ dependences (dark solid and pale
dotted curves, respectively) in a broad range of $K$. Note the
periodic appearance of patterns of similar structure.}
\label{fig5}
\end{figure}

\end{document}